# MODELING OF AN INDUCTIVE ADDER KICKER PULSER FOR DARHT-II[*]

L. Wang, G. J. Caporaso, E. G. Cook, LLNL, Livermore, CA94550, USA

*Abstract*

An all solid-state kicker pulser for a high current induction accelerator (the Dual-Axis Radiographic Hydrodynamic Test facility DARHT-2) has been designed and fabricated. This kicker pulser uses multiple solid state modulators stacked in an inductive-adder configuration. Each modulator is comprised of multiple metal-oxide-semiconductor field-effect transistors (MOSFETs) which quickly switch the energy storage capacitors across a magnetic induction core. Metglas is used as the core material to minimize loss. Voltage from each modulator is inductively added by a voltage summing stalk and delivered to a 50 ohm output cable. A lumped element circuit model of the inductive adder has been developed to optimize the performance of the pulser. Results for several stalk geometries will be compared with experimental data.

## 1 INTRODUCTION

Linear induction accelerator based x-ray technology can provide time-resolved, 3-D radiography capabilities for a hydrodynamic event. A key component of this technology is a kicker system[1]. The kicker system cleaves a series of intense electron beam micropulses, and steers the beam into separate beam transport lines to achieve multiple lines of sight. The first part of this fast kicker system is a high current stripline dipole kicker. The original design of the pulser which drives this stripline kicker was based on planar triodes[2]. Although the performance of the pulser based on this design was very good, the availability of the high frequency planar triodes in the future has become a concern. This led to the development of an all solid-state kicker pulser design for the Dual-Axis Radiographic Hydrodynamic Test facility DARHT-2. The new pulser design was based on the Advanced Radiograph Machine (ARM) modulator technology [3]. It uses multiple solid-state modulators stacked in an inductive-adder configuration. Each modulator is comprised of multiple metal-oxide-semiconductor field-effect transistors (MOSFETs) which quickly switch the energy storage capacitors across a magnetic induction core. Metglas is used as the core material to minimize loss. Voltage from each modulator is inductively added by a voltage summing stalk and delivered to a 50 ohm output cable. The cross section of this solid-state kicker pulser is shown in Figure 1. There are thirty stacked modules in the kicker pulser. For illustration purpose, only three stacked modules are shown in the figure. On each stack, there are 24 MOSFETs. The secondary winding is a metal rod which is placed on the centerline of the kicker pulser. The rod can be grounded at either end to generate a positive or a negative output voltage. A picture of the fabricated solid-state kicker pulser is shown in Figure 2.

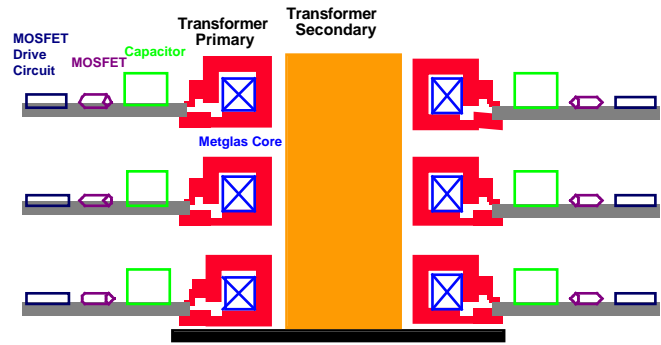

Fig. 1: Cross section of the solid-state kicker pulser.

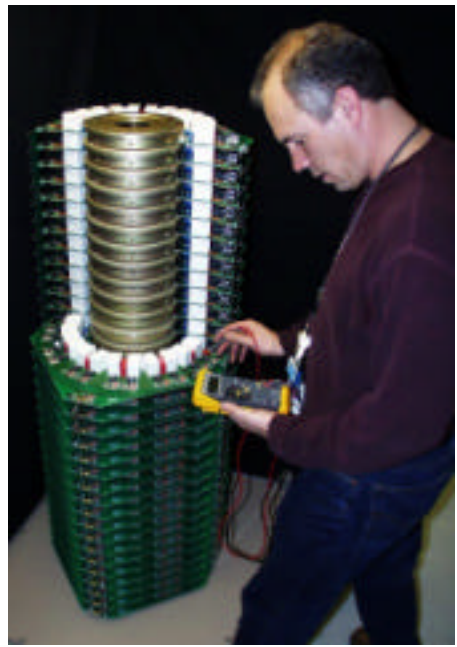

Figure 2: Inductive adder kicker pulser.

---

## 2 INDUCTIVE ADDER MODEL

A lumped element circuit model of the inductive adder has been developed to simulate the output voltage waveform and to determine the choice of stalk geometry. Figure 3 shows the lumped element equivalent circuit model of the inductive adder.

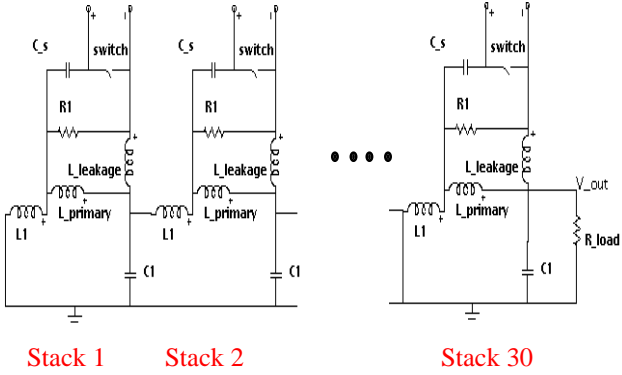

Figure 3: A lumped element equivalent circuit model of the inductive adder.

Thirty stacks are simulated. Each stack is modeled with a subcircuit which consists of a switch, capacitors, inductors, and a resistor. The values of these components are determined from the detailed circuit diagram of each module. The capacitance C_s and the resistance R1 are 24 µF and 50 Ohms, respectively. The load resistance R_load is 50 Ohms. The inductance of the transformer primary L_primary is 20.9 µH. The effective leakage inductance L_leakage is 20 nH. The inductance L1 and capacitance C1 are determined from $a$, the radius of the stalk under consideration, and $b$, the inner wall radius of kicker pulser using the following equations:

$$L1 = \frac{\mu_0 h \ln \frac{b}{a}}{2\pi} \quad (1)$$

$$C1 = \frac{2\pi \varepsilon_0 h}{\ln \frac{b}{a}} \quad (2)$$

where $h$ is the height of each stack. The stalk impedance is

$$Z = \sqrt{\frac{L1}{C1}} \quad (3)$$

## 3 SIMULATION RESULTS

Output voltage waveforms are simulated using the circuit model for different stalk geometries. These cases are chosen based on the stalks made for the experiment. Figure 4 shows the plots of output voltage versus time with a 50 Ohm stalk. The capacitors are charged to 650 V initially. With thirty modulator stacks, the total voltage should go to 19.5 kilovolts. All switches are triggered simultaneously. For the simulation, the switches are closed from 10 ns to 100 ns. With a 50 Ohm stalk, L1 and C1 are 6.5 nH and 2.6 pF, respectively. If a stalk is matched to the load, we will expect the output voltage waveform to be a square wave except during the rise and fall time. The output voltage plot of the 30th stack in Figure 4 shows that the 50 Ohm stalk doesn't match into the 50 Ohm load. The waveform indicates that the impedance of the stalk should be reduced in order to match into the load impedance. This is due to the presence of the effective leakage inductance. The plots of output voltage versus time with a 12.6 Ohm stalk are shown in Figure 5. With a 12.6 Ohm stalk, L1 and C1 are 1.6 nH and 10.3 pF, respectively. There is an overshoot in the output voltage waveform in Figure 5. This indicates that in order to match into the load impedance, the impedance of the stalk should be increased. The plots of output voltage versus time with a 18.9 Ohm stalk are displayed in Figure 6. With a 18.9 Ohm stalk, L1 and C1 are 2.5 nH and 6.9 pF, respectively. The output voltage waveform shows that this 18.9 Ohm stalk provides a better match into the load impedance compared to other available stalks. This result agrees with the experimental data.

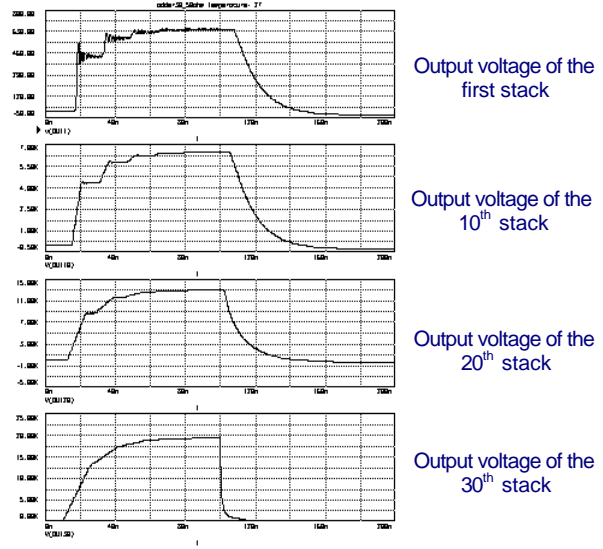

Figure 4: Output voltage versus time with a 50 Ω stalk.

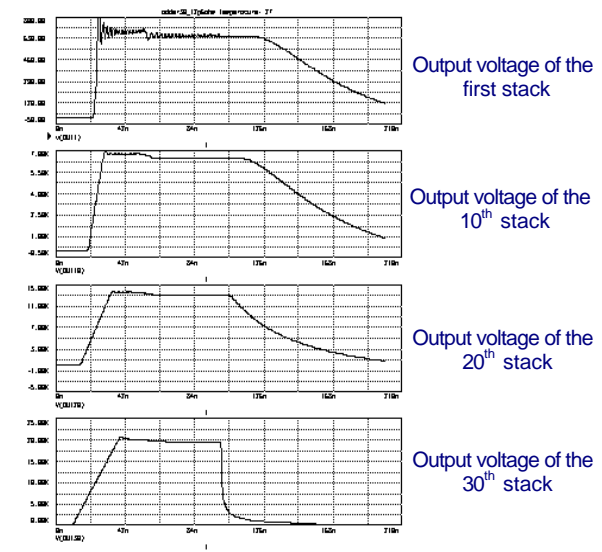

Figure 5: Output voltage versus time with a 12.6 Ω stalk.

The output voltage versus time from experimental data with a 50 Ohm stalk is shown in Figure 7. For the experiment, the output voltage has a negative value. We can see that the output voltage waveforms are similar (except longer rise time) between the simulation result and experimental data. The output voltage versus time from the experimental data with a 12.5 Ohm stalk is shown in Figure 8. There is an overshoot in the output voltage waveform as we have also seen from the simulation result.

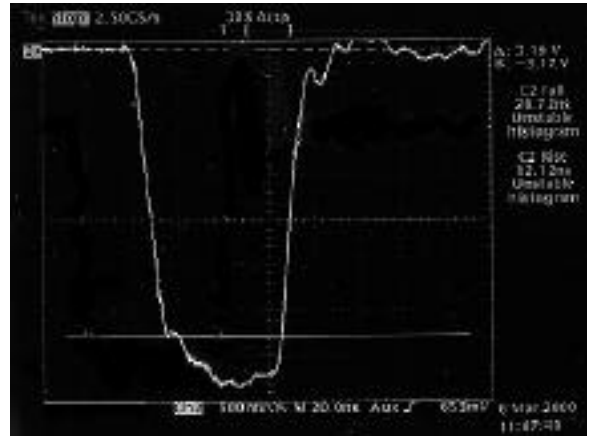

Figure 7: Output voltage versus time from experimental data (with a 50 stalk).

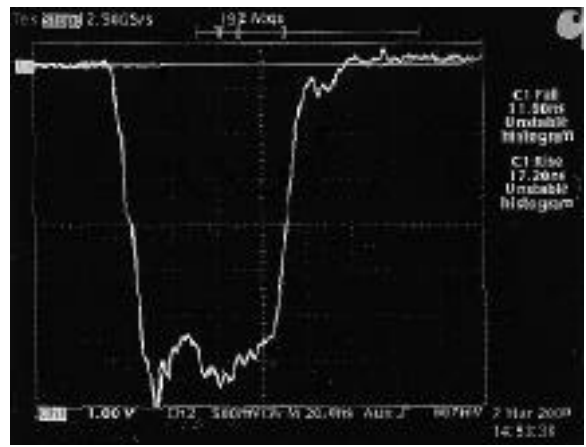

Figure 8: Output voltage versus time from experimental data (with a 12.5 stalk).

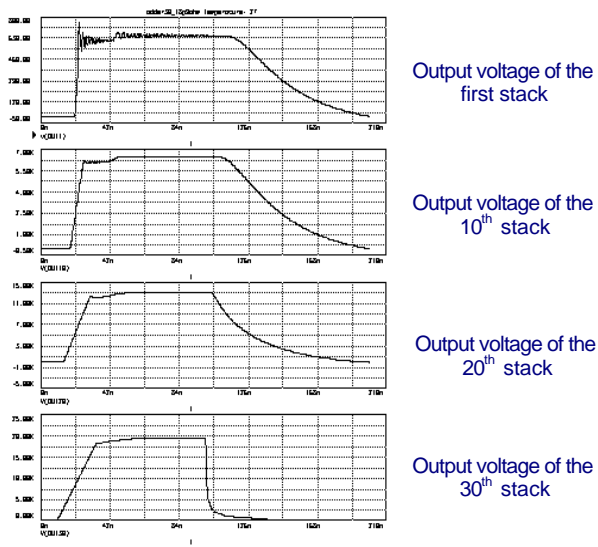

Figure 6: Output voltage versus time with a 18.9 stalk.

## 4 SUMMARY

A lumped element circuit model of the inductive adder kicker pulser has been developed. Output voltage waveforms are simulated using the circuit model for different stalk geometries. Based on the estimated effective leakage inductance, the simulation results agree with the experimental data on the choice of stalk.